\documentclass[twocolumn,nofootinbib]{revtex4}
\usepackage{graphicx}
\usepackage{latexsym}

\def\be{\begin{equation}}
\def\ee{\end{equation}}
\def\bea{\begin{eqnarray}}
\def\eea{\end{eqnarray}}
\newcommand{\DE}{{\mathrm{de}}}
\newcommand{\DM}{{\mathrm{dm}}}

\begin{document}

\title{Constraining Cosmological Parameters with Observational Data \\ Including Weak Lensing Effects}

\author{Hong Li${}^{a,b}$}
\author{Jie Liu${}^{a}$}
\author{Jun-Qing Xia${}^{c}$}
\author{Lei Sun${}^{d}$}
\author{Zu-Hui Fan${}^{d}$}
\author{Charling Tao${}^e$}
\author{Andre Tilquin${}^e$}
\author{Xinmin Zhang${}^{a,b}$}

\affiliation{${}^a$Institute of High Energy Physics, Chinese
Academy of Science, P.O.Box 918-4, Beijing 100049, P.R.China}

\affiliation{${}^b$Theoretical Physics Center for Science Facilities
(TPCSF), Chinese Academy of Science, P.R.China}

\affiliation{${}^c$Scuola Internazionale Superiore di Studi
Avanzati, Via Beirut 2-4, I-34014 Trieste, Italy}

\affiliation{${}^d$Department of Astronomy, School of Physics,
Peking University, Beijing 100871, P.R.China}

\affiliation{${}^e$Centre de Physique des Particules de Marseille,
CNRS/IN2P3-Luminy and Universit\'e de la M\'editerran\'ee, Case 907,
F-13288 Marseille Cedex 9, France}

%\affiliation{${}^d$Department of Physics, Simon Fraser University,
%Burnaby, BC V5A 1S6, Canada}

%\date{\today}

\begin{abstract}

In this paper, we study the cosmological implications of the 100
square degree Weak Lensing survey (the CFHTLS-Wide, RCS,
VIRMOS-DESCART and GaBoDS surveys). We combine these weak lensing
data with the cosmic microwave background (CMB) measurements from
the WMAP5, BOOMERanG, CBI, VSA, ACBAR, the SDSS LRG matter power
spectrum and the Type Ia Supernoave (SNIa) data with the ``Union"
compilation (307 sample), using the Markov Chain Monte Carlo method
to determine the cosmological parameters, such as the
equation-of-state (EoS) of dark energy $w$, the density fluctuation
amplitude $\sigma_8$, the total neutrino mass $\sum m_{\nu}$ and the
parameters associated with the power spectrum of the primordial
fluctuations. Our results show that the $\Lambda$CDM model remains a
good fit to all of these data. In a flat universe, we obtain a tight
limit on the constant EoS of dark energy, $w=-0.97\pm 0.041$
($1\sigma$). For the dynamical dark energy model with time evolving
EoS parameterized as $w_{\DE}(a) = w_0 + w_a (1-a)$, we find that
the best-fit values are $w_0=-1.064$ and $w_a=0.375$, implying the
mildly preference of Quintom model whose EoS gets across the
cosmological constant boundary during evolution. Regarding the total
neutrino mass limit, we obtain the upper limit, $\sum m_{\nu}<
0.471$ eV ($95\%$ C.L.) within the framework of the flat
$\Lambda$CDM model. Due to the obvious degeneracies between the
neutrino mass and the EoS of dark energy model, this upper limit
will be relaxed by a factor of $2$ in the framework of dynamical
dark energy models. Assuming that the primordial fluctuations are
adiabatic with a power law spectrum, within the $\Lambda$CDM model,
we find that the upper limit on the ratio of the tensor to scalar is
$r<0.35$ ($95\%$ C.L.) and the inflationary models with the slope
$n_s\geq1$ are excluded at more than $2~\sigma$ confidence level. In
this paper we pay particular attention to the contribution from the
weak lensing data and find that the current weak lensing data do
improve the constraints on matter density $\Omega_m$, $\sigma_8$,
$\sum{m_{\nu}}$, and the EoS of dark energy.
\end{abstract}

%\pacs{98.80.Es; 98.80.Cq}

\maketitle

%Introduction==========================================================

\section{Introduction}
\label{Int}

Cosmological observations play a crucial role in understanding our
universe. With the accumulation of observational data from CMB
measurements, large scale structure (LSS) surveys and SNIa
observations as well as the improvements of the data quality, the
accurate constraints on cosmological parameters can be expected from
the data analysis. Recently WMAP group has released the five-year
data of temperature and polarization power spectra
\cite{WMAP5GF1,WMAP5GF2,WMAP5Other}. And the Arcminute Cosmology
Bolometer Array (ACBAR) experiment has also published its new CMB
temperature power spectrum \cite{ACBAR}. These new CMB data can
strengthen the constraints on the cosmological parameters,
especially for those associated with the inflationary models
\cite{KinneyWMAP5,WMAP5GF1,WMAP5GF2}. Furthermore, the Supernova
Cosmology Project has made an unified analysis of the world's
supernovae datasets and presented a new compilation ``Union" (307
sample) \cite{Union} which includes the recent samples of SNIa from
SNLS and ESSENCE Survey, as well as some older datasets. In the
literature \cite{Union,LinderUnion,xia_union}, these data has been
widely used to constrain various cosmological models. However, one
should keep in mind that the degeneracies of cosmological parameters
exist in almost all cosmological observations, \emph{i.e.}, they are
not sensitive to single parameters but to some specific combinations
of them. It is therefore highly necessary to combine different
probes to break parameter degeneracies so as to achieve tight
constraints. Furthermore, different observations are affected by
different systematic errors, and it is thus helpful to reduce
potential biases by combining different probes.

Weak gravitational lensing which is directly related to the dark
matter distribution, the geometry and the dynamics of the universe
provides an useful way to break the degeneracies mentioned above.
Cosmic shear, i.e. the distortion of images of the high-redshift
source galaxies through the tidal gravitational field of the
large-scale matter distribution in the universe,  provides a
powerful probe of the mass fluctuations in the universe and gives a
direct measurement of the matter power spectrum down to the
non-linear regime. Since the mass distribution and its late time
evolution are sensitive to the dark energy and the neutrino mass, we
expect to get interesting information about both through the study
of cosmic shear.

Recently, the cosmic shear analysis of the 100 square degree weak
lensing survey which includes the Canada-France-Hawaii Telescope
Legacy Survey (CFHTLSWide)\cite{Hoekstra2006}, the Garching-Bonn
Deep Survey (GaBoDs)\cite{Hetterscheidt2006}, the Red-Sequence
Cluster Survey (RCS)\cite{Hoekstra2002a}, and the VIRMOS-DESCART
survey (VIRMOS)\cite{Waerbeke2005} has been presented by
Ref.\cite{Benjamin2007}. These combined surveys cover $113~{\rm
degree}^{2}$ which is the largest sky coverage so far. With these
data, there are some studies on the cosmological implications
recently in the literature
\cite{ichiki2008,Gongyan2008,Lesgourgues2007,Shaun2008,FuLiping2008,Fu2008,Kilbinger2008}.
% Cosmic shear
%is a very promising technique for studying the large-scale
%structure, it provides a potentially powerful probe of the mass
%fluctuations in the Universe and gives a direct measurement of the
%matter power spectrum to non-linear regime. The mass distribution,
%and its late time evolution, are sensitive to both the dark energy,
%the mystic power to drive universe's acceleration, the neutrino mass
%and the cosmological parameters which relate to it.
In this paper we study the constraints on cosmological parameters
from the weak lensing cosmic shear as well as the current
cosmological observational data, such as CMB, LSS and SN Ia.
Specifically, we focus on the effects from the weak lensing.

Our paper is organized as follows: In Section II we describe the
method and the latest observational datasets we use in this paper;
Section III contains our main global fitting results on the
cosmological parameters and the last section is the summary.

%Method and Current Observations=======================================

\section{Method and Data}
\label{Method}

In our study, we perform a global analysis using the publicly
available MCMC package CosmoMC\footnote{Available at:
http://cosmologist.info/cosmomc/.} \cite{CosmoMC}. We assume the
purely adiabatic initial conditions. Our most general parameter
space is:
\begin{equation}
\label{parameter} {\bf P} \equiv (\omega_{b}, \omega_{c},
\Theta_{s}, \tau, w_{0}, w_{a}, f_{\nu}, n_{s}, A_{s}, \alpha_s,
r)~,
\end{equation}
where $\omega_{b}\equiv\Omega_{b}h^{2}$ and
$\omega_{c}\equiv\Omega_{c}h^{2}$, in which $\Omega_{b}$ and
$\Omega_{c}$ are the physical baryon and cold dark matter densities
relative to the critical density, $\Theta_{s}$ is the ratio
(multiplied by 100) of the sound horizon to the angular diameter
distance at decoupling, $\tau$ is the optical depth to
re-ionization, $f_{\nu}$ is the dark matter neutrino fraction at
present, namely,
\begin{equation}
f_{\nu}\equiv\frac{\rho_{\nu}}{\rho_{\DM}}=\frac{\Sigma
m_{\nu}}{93.105~\mathrm{eV}~\Omega_ch^2}~.
\end{equation}
The primordial scalar power spectrum $\mathcal{P}_{\chi}(k)$ is
parameterized as \cite{Ps}:
\begin{eqnarray}
\ln\mathcal{P}_{\chi}(k)=\ln
A_s(k_{s0})&+&(n_s(k_{s0})-1)\ln\left(\frac{k}{k_{s0}}\right)\nonumber\\
&+&\frac{\alpha_s}{2}\left(\ln\left(\frac{k}{k_{s0}}\right)\right)^2,
\end{eqnarray}
where $A_s$ is defined as the amplitude of initial power spectrum,
$n_s$ measures the spectral index, $\alpha_{s}$ is the running of
the scalar spectral index and $r$ is the tensor to scalar ratio of
the primordial spectrum. For the pivot scale we set
$k_{s0}=0.05$Mpc$^{-1}$. Moreover, $w_0$ and $w_a$ are the
parameters of dark energy EoS, which is given by \cite{Linderpara}:
\begin{equation}
\label{EOS} w_\DE(a) = w_{0} + w_{a}(1-a)~,
\end{equation}
where $a=1/(1+z)$ is the scale factor and $w_{a}=-dw/da$
characterizes the ``running" of EoS (RunW henceforth). The
$\Lambda$CDM model has $w_0=-1$ and $w_a=0$. For the dark energy
model with a constant EoS, $w_a=0$ (WCDM henceforth). When using the
global fitting strategy to constrain the cosmological parameters, it
is crucial to include dark energy perturbations
\cite{WMAP3GF,LewisPert,XiaPert}. In this paper we use the method
provided in Refs.\cite{XiaPert,ZhaoPert} to treat the dark energy
perturbations consistently in the whole parameter space in the
numerical calculations.

For the weak lensing likelihood statistic, we mainly consider the
shear correlation function $\xi$ following Ref. \cite{Benjamin2007}.
The shear correlation functions are defined as
\begin{eqnarray}
\xi_+(\theta)&=& \langle
\gamma_{t}(r)\gamma_t(r+\theta)\rangle+\langle
\gamma_r(r)\gamma_r(r+\theta)
\rangle.\nonumber \\
\xi_-(\theta)&=&\langle \gamma_t(r)\gamma_t(r+\theta)\rangle
-\langle \gamma_r(r)\gamma_r(r+\theta)\rangle, \label{xipm}
\end{eqnarray}
where the shear $\Upsilon = (\gamma_t, \gamma_r)$ is rotated into
the local frame of the line joining the centres of each galaxy pair
separated by $\theta$.  The shear correlation function $\xi_+$ is
related to the convergence power spectrum through
\begin{equation}
\xi_+(\theta)={1\over 2\pi} \int_0^\infty~{\rm d} k~
 k P_\kappa(k) J_0(k\theta),
\label{theogg}
\end{equation}
\noindent where $J_0$ is the zeroth order Bessel function, and
$P_\kappa(k)$ is the power spectrum of the projected density field,
as given by
\begin{eqnarray}
P_\kappa(k)&=&\frac{9\,\rm{H}_0^4\,\Omega_{\mathrm{m}}^2}{4c^4}\int_0^{r_H}
{{\rm d}r \over a^2(r)} P_{\delta}\left({k\over f_K(r)};
r\right)\times\nonumber\\
&&\left[ \int_r^{r_H}{\rm d} r' n(r') {f_K(r'-r)\over
f_K(r')}\right]^2, \label{pofkappa}
\end{eqnarray}
\noindent where $\rm{H}_0$ is the Hubble constant, $f_K(r)$ is the
comoving angular diameter distance out to a distance $r$ ($r_H$ is
the comoving horizon distance), $a(r)$ is the scale factor, and
$n[r(z)]$ is the redshift distribution of the sources\footnote{In
our calculations, we use the redshift distribution given in the EQ
(11) of Ref.\cite{Benjamin2007} and the best fit values of the
nuisance parameters a,b and c presented in their Table 2. We have
numerically checked that the correlations coefficients between a,b,c
parameters and cosmological parameters are small enough to be
neglected. So the main global analysis results presented in this
paper are obtained with the fixed nuisance parameters. Similar
conclusions about varying a,b and c are also given in Ref.
\cite{ichiki2008}.}. $P_{\delta}$ is the 3-dimensional mass power
spectrum computed from a non-linear estimation of the dark matter
clustering, and $k$ is the 2-dimensional wave vector perpendicular
to the line-of-sight. $P_{\delta}$ evolves with time, hence depends
on the co-moving radial coordinate $r$.

The lensing signal can be split into E mode and B mode, and the E
and B shear correlation functions are given by
\begin{equation}
\xi_E(\theta)=\frac{\xi_+(\theta)+\xi'(\theta)}{2},\ \ \ \ \ \
\xi_B(\theta)=\frac{\xi_+(\theta)-\xi'(\theta)}{2}, \label{eqn:xieb}
\end{equation}
where \begin{equation}
\xi'(\theta)=\xi_-(\theta)+4\int_\theta^\infty \frac{{\rm
d}\vartheta}{\vartheta} \xi_-(\vartheta)
    -12\theta^2 \int_\theta^\infty \frac{{\rm d}\vartheta}{\vartheta^3}\xi_-(\vartheta).
\label{eqn:xipr}
\end{equation}

In this paper we use the E mode correlation as the data vector. The corresponding
likelihood function is given by
\begin{equation}
{\cal L}={1\over \sqrt{(2\pi)^n|C|}}
\exp\left[-\frac{1}{2}(\xi_E-m)C^{-1}(\xi_E-m)^T\right],
\label{likelihood}
\end{equation}
where $n$ is the number of angular scale bins and $C$ is the
$n\times n$ shear covariance matrix. We take the covariance matrix
given by Ref. \cite{Benjamin2007} which includes the statistical
noise, the residual B-mode and the sample variance.

In the computation of the CMB, we include the WMAP5 temperature and
polarization power spectra with the routine for computing the
likelihood supplied by the WMAP team\footnote{Available at the
LAMBDA website: http://lambda.gsfc.nasa.gov/.}. We also include some
small-scale CMB measurements, such as BOOMERanG \cite{BOOMERanG},
CBI \cite{CBI}, VSA \cite{VSA} and the newly released ACBAR data
\cite{ACBAR}. For the Large Scale Structure information, we use the
Sloan Digital Sky Survey (SDSS) luminous red galaxy (LRG) sample
\cite{Tegmark:2006az}. The supernova data we use are the recently
released ``Union" compilation of 307 sample \cite{Union}. In the
calculation of the likelihood from SNIa we marginalize over the
relevant nuisance parameter \cite{SNMethod}.

Furthermore, we make use of the Hubble Space Telescope (HST)
measurement of the Hubble parameter $H_{0}\equiv
100$h~km~s$^{-1}$~Mpc$^{-1}$ by a Gaussian likelihood function
centered around $h=0.72$ and with a standard deviation $\sigma=0.08$
\cite{HST}.

%Results===============================================================

\section{Numerical Results}
%Results(Dark Energy)==================================================

In this section we present our global fitting results. The most
relevant cosmological parameters for the weak gravitational lensing
signal are the matter density $\Omega_m$ and the matter power
spectrum normalization parameter $\sigma_8$. On the other hand,
other cosmological parameters related to the primordial power spectrum from inflation
and the expansion history of the universe can also affect the weak lensing effects.
Thus we expect that weak lensing data can provide important constraints on these
parameters as well.
%As we know that, the
%measured cosmic shear signals come from the galaxy shape catalogs,
%and it is directly related to the two-point statistics function. The
%observed two-point statistics can be related to the matter power
%spectrum $P(k)$, which depends on the background cosmological
%parameters and the primordial power spectrum from the inflation. For
%the cosmic shear, some of the dependence on cosmology also appear
%through the angular diameter distances to the sources. So, the
%constraints can be expected from the cosmic shear signal.

\begin{figure}[t]
\begin{center}
\includegraphics[scale=0.5]{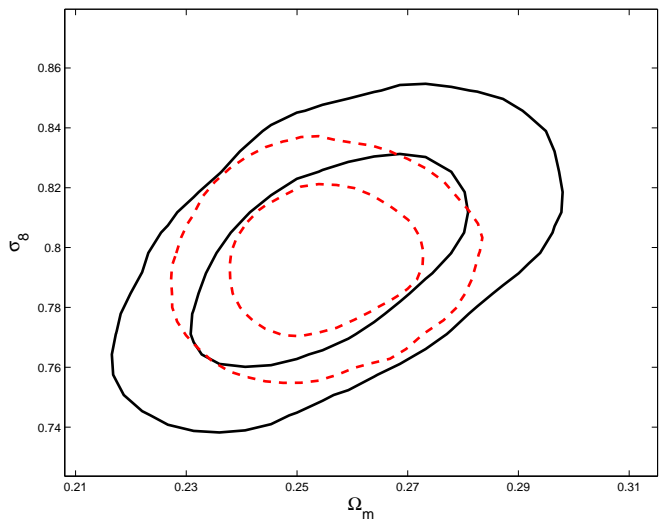}
\caption{Two dimensional constraints on $\sigma_{8}$ and $\Omega_m$
from the current observations in $\Lambda$CDM model, assuming a flat
universe. The black solid lines are from data sets CMB $+$ LSS $+$
SN Ia and the red dashed lines are obtained by taking into account
the weak lensing data. \label{fig5}}
\end{center}
\end{figure}

\begin{table}\hspace{-5mm}
TABLE I. Constraints on the parameters $n_s$, $\alpha_s$,
$r$ and the total neutrino mass $\sum m_{\nu}$ from the data
combination with or without the weak lensing effects. We have shown the mean
$1~\sigma$ errors. For the weakly constrained parameters we quote
the $95\%$ upper limits instead.
\begin{center}\hspace{-15mm}

\begin{tabular}{|c|c|c|c|}

\hline

& \multicolumn{2}{|c|}{$\Lambda$CDM} & RunW \\

\cline{2-4}

& with wl & w/o wl& with wl\\

\hline

$\sum m_{\nu}$&$<0.471 eV(95\%)$&$<0.663 eV(95\%)$&$<1.110 eV(95\%)$ \\

\hline

$n_s$&$0.953\pm0.0202$&$0.946\pm0.0217$& $...$\\

\hline

$\alpha_s$&$-0.0231\pm0.0185$&$-0.0281\pm0.0202$&$...$\\

\hline

$r$&$<0.346$ ($95\% $)&$<0.386$ ($95\%$)&$...$\\

\hline
\end{tabular}
\end{center}
\end{table}

First of all, we consider the flat $\Lambda$CDM model. To see the
effect of the current weak lensing data, we calculate and compare
the results for the two cases with and without the lensing data
included. In Figure \ref{fig5} we present the constraints on
$\Omega_m$ and $\sigma_8$. A significant improvement from the cosmic
shear data can be seen clearly. We obtain $\Omega_m=0.25\pm0.011$
and $\sigma_8=0.80\pm0.016$ at $1 \sigma$ C.L. from the combined
CMB, LSS, SN and weak lensing cosmic shear data in $\Lambda$CDM
model. Compared with the results with no weak lensing data included,
the constraints are improved by $45\%$ and $40\%$ for $\Omega_m$ and
$\sigma_8$, respectively.

In the following, we will present the constraints on the neutrino
mass, the equation of state of dark energy and the
parameters associated with inflation models respectively.

\subsection{Neutrino Mass} \label{Mnu}

From the neutrino oscillation experiments, such as the atmospheric
neutrino experiments \cite{Atmospheric} and the solar neutrino
experiments \cite{Solar}, we know that neutrinos are massive.
Therefore they can
%Neutrinos, as a mysterious budget of our universe, can
leave some unique imprints on cosmological observables, such as
the CMB anisotropies and LSS \cite{NeuRev}.
%temperature power spectrum and the LSS matter power spectrum \cite{NeuRev},
Thus cosmological observations can provide crucial complementary information on the absolute
neutrino masses.

\begin{figure}[t]
\begin{center}
\includegraphics[scale=0.4]{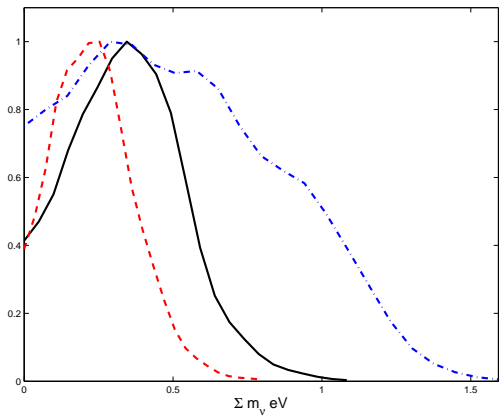}
\caption{One dimensional probability distribution of the total
neutrino mass $\Sigma {m_{\nu}}$from the current observations
assuming a flat universe. The black solid line is from data sets CMB
$+$ LSS $+$ SN Ia, the red dashed line is obtained by taking into
account the weak lensing data and the blue dash-dotted line is for
the dynamical dark energy model. \label{fig3}}
\end{center}
\end{figure}

Besides the influence on the expansion history of the universe, the existence
of massive neutrinos affect the structure formation process.
%Cosmological observations potentially provides a direct weighing of
%the neutrino mass through the neutrino's kinematic effects on large
%scale structure formation.
Massive neutrinos contribute to the dark matter composition of the
universe.
%at late time of the universe,
Because of their relatively large thermal velocities, they damp the
perturbations within their free streaming scale, and suppress the
matter power spectrum at small scale by roughly
$\Delta{P}/P\sim-8\Omega_{\nu}/\Omega_m$ \cite{Suppress}. Such a
suppression of the power spectrum depends on the Jeans length of
neutrinos, which decreases with time as the neutrino thermal
velocity decreases. Since the cosmic shear can measure the matter
power spectrum down to small-scale non-linear regime, it can be used
to significantly improve the constraint on the neutrino mass.

Within the $\Lambda$CDM model, from Table I one can read out $95\%$
upper limit of the total neutrino mass derived from the current
observations, CMB+LSS+SN, $\sum m_{\nu}<0.663~\mathrm{eV}$ ($95\%$
C.L.), which is consistent with the recent results from WMAP5 group
\cite{WMAP5GF1,WMAP5GF2}. By taking into account the weak lensing
cosmic shear data, the constraints can be tightened to
$\sum m_{\nu}<0.471~\mathrm{eV}$ ($95\%$ C.L.).

The degeneracies between the neutrino mass and
other cosmological parameters, such as the EoS of dark
energy \cite{Hannestad} and the running of spectral index
\cite{NeuRun}, are known to exist. We thus also consider the
neutrino mass limit in the framework of the dynamical dark energy
model. One can see from the blue curve in Figure \ref{fig3} that
the degeneracy\cite{WMAP3GF,Hannestad,XiaMnu}
causes a much looser limit on the neutrino mass with
$\sum m_{\nu}<1.11~\mathrm{eV}$ ($95\%$ C.L.).

\begin{figure}[t]
\begin{center}
\includegraphics[scale=0.4]{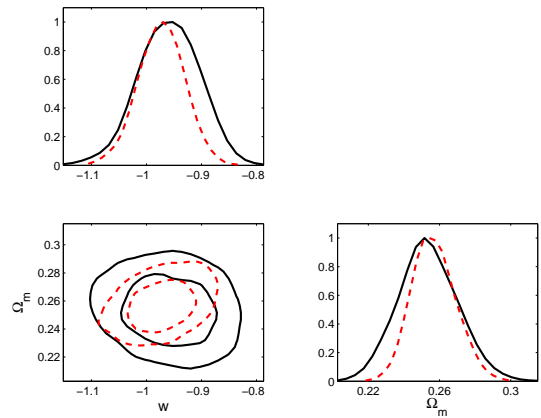}
\vspace{9mm} \caption{One dimensional and two dimensional cross
correlation constraints on the constant EoS of dark energy, $w$, and
the present matter density, $\Omega_m$, from the combined data,
assuming a flat universe. The black solid lines are from data sets
CMB $+$ LSS $+$ SN Ia and the red dashed lines are obtained by
taking into account the weak lensing data. \label{fig1}}
\end{center}
\end{figure}

\begin{figure}[t]
\begin{center}
\includegraphics[scale=0.4]{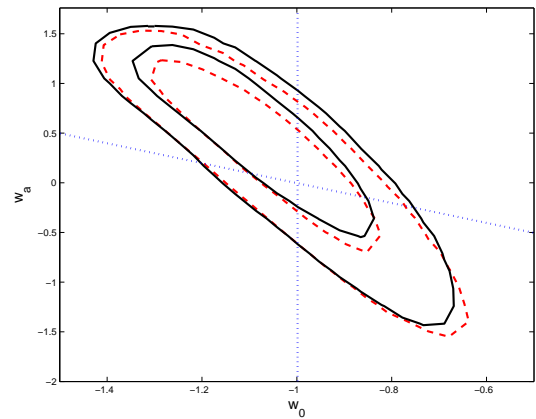}
\vspace{2mm} \caption{Two dimensional constraints on the dark energy
EoS parameters $w_0$ and $w_a$ from the current observations,
assuming a flat universe. The black solid lines are from data sets
CMB $+$ LSS $+$ SN Ia and red dashed lines are obtained by taking
into account the weak lensing data.  The blue dashed lines stand for
$w_0=-1$ and $w_0+w_a=-1$.\label{fig2}}
\end{center}
\end{figure}

\subsection{Equation of State of Dark Energy}
\label{DEEoS}

\begin{table*}{\footnotesize
TABLE II. Constraints on the dark energy EoS and some background
parameters from the observations. Here we have presented the mean
and the best fit values, which are obtained from the cases with and
without the weak lensing data, respectively.
\begin{center}

\begin{tabular}{|c|c|c|c|c|c|c|c|c|c|c|c|}

\hline

\multicolumn{2}{|c|}{Parameter}&\multicolumn{2}{c|}{$w_0$}&\multicolumn{2}{c|}{$w_{a}$}&\multicolumn{2}{c|}{$\Omega_\DE$}&\multicolumn{2}{c|}{$H_0$}\\

\cline{3-10}

\multicolumn{2}{|c|}{}& with wl & w/o wl & with wl & w/o wl &
with wl & w/o wl & with wl & w/o wl \\

\hline

$\Lambda$CDM & BestFit & $-1$ & $-1$ & $0$ & $0$ & $0.742$ & $0.749$ & $71.63$ & $72.07$\\

\cline{2-10}

flat& Mean & $-1$ & $-1$ & $0$ & $0$ & $0.746\pm0.0109$ & $0.744\pm0.0160$ & $71.98\pm1.181$ & $71.87\pm1.511$\\

\hline

WCDM & BestFit & $-0.967$ & $-0.992$ & $0$ & $0$ & $0.741$ & $0.742$ & $71.15$ & $71.53$\\

\cline{2-10}

flat& Mean & $-0.973\pm0.0413$ & $-0.959\pm0.0544$ & $0$ & $0$ & $0.743\pm0.012$ & $0.746\pm0.0163$ & $71.48\pm1.46$ & $71.53\pm1.58$\\

\hline

RunW & BestFit & $-1.064$ & $-1.066$ & $0.375$ & $0.514$ & $0.737$ & $0.738$ & $70.74$ & $70.01$\\

\cline{2-10}

flat& Mean & $-1.035\pm0.157$ & $-1.059\pm0.160$ & $0.207\pm0.644$ & $0.360\pm0.657$ & $0.737\pm0.0170$ & $0.738\pm0.0193$ & $70.76\pm2.080$& $70.46\pm2.163$ \\

\hline

%RunW & BestFit & $-1.095$ & $...$ & $0.553$ & $...$ & $0.729$ & $...$ & $71.27$ & $...$\\

%\cline{2-10}

%non-flat& Mean & $-1.029\pm0.157$ & $...$ & $0.197\pm0.636$ & $...$ & $0.737\pm0.0170$ & $...$ & $70.92\pm2.812$& $...$ \\

\hline

\end{tabular}
\end{center}}
\end{table*}

For the dark energy component, in Table II we list the constraints
on the EoS, dark energy density and the Hubble constant in different
dark energy models. In order to emphasize the contribution from the
cosmic shear measurements, we compare the following two cases:
combining CMB, LSS and SN Ia with or without weak lensing data. We
find that in both cases the $\Lambda$CDM model remains a good fit to
the data.

Considering dynamical dark energy models in flat universe, we first explore the constraints on
the constant EoS of dark energy, $w$. In Fig.\ref{fig1} we show the
one dimensional probability distributions and the two dimensional
cross correlation of $w$ and the present dark matter density,
$\Omega_m$. One can find that the weak lensing measurements provide a significant
improvement, notably seen in the two-dimensional constraints on $\Omega_m$ and $w$.
The current observations including the weak lensing data yield a strong constraint on the constant
EoS of dark energy, $w=-0.973\pm0.0413$ ($1~\sigma$).
This result is similar to and somewhat tighter than the limit from WMAP5 \cite{WMAP5GF1},
$w=-0.972^{+0.061}_{-0.060}$ ($1~\sigma$). It excludes some
of the quintessence models, for example, the tracker model which
predicts $w\sim-0.7$ \cite{Tracking} at $5~\sigma$.

We also consider the time evolving EoS, parameterized as
$w_\DE(a)=w_0+w_{a}(1-a)$. In Fig.\ref{fig2} we illustrate the
constraints on $w_0$ and $w_{a}$. For the flat universe, combining
all the data,  we obtain the $2\sigma$ limit
$w_0=-1.035^{+0.324}_{-0.274}$ and $w_a=0.207^{+0.991}_{-1.439}$.
The best fit values are $w_0=-1.064$ and $w_{a}=0.375$, which show
that $w(z)$ crosses the cosmological constant boundary during the
evolution described by the Quintom dark energy
model\cite{Quintom,zhangxiaofei2005}. The large variances of $w_0$
and $w_{a}$ with $-1.309<w_0<-0.711$ and $-1.232<w_{a}<1.198$ at
$2\sigma$, show that at current stage, the allowed parameter space
for $w_0$ and $w_a$ is still large. On the other hand, the
$\Lambda$CDM model remains an excellent fit. Comparing the two fits
with and without weak lensing data, we find that the current cosmic
shear data do not contribute significantly to the constraints on the
varying equation of state of dark energy due to the large
statistical errors in the data.
%The main reason for this is that the mean redshift of the galaxies is
%around $z \sim 0.78$, which overlaps with the redshift scale of the
%SN Ia which we use, further more there exists large errors of these
%data at current stage.
We expect that future wide and deep weak lensing observations will
produce data with dramatically reduced statistical errors, and thus
provide significant improvements in constraining dark energy.
However, different systematics involved in weak lensing effects have
to be thoroughly investigated in order to fully realize its
constraining power.

%Results(Others)=======================================================
%\subsection{Other Cosmological Parameters}
%Results(Others,Curvature)=============================================
%Results(Others,Neutrino)==============================================
%Results(Others,Inflation)=============================================

\subsection{Parameters of the primordial fluctuations} \label{Inf}

In this section we present the constraints on the
parameters related to the primordial fluctuations, which are closely
associated with inflationary models.
%Since the quantum fluctuations of the inflation field
%turn out to be the primordial density fluctuations which seed the
%observed large scale structures, the cosmic shear can provide the
%information of the early universe.
%Currently, the observational data are in good agreement with models that
%generate a gaussian, adiabatic
%and nearly scale-invariant primordial spectrum, which is consistent
%with single field slow-roll inflation predictions.

\begin{figure}[t]
\begin{center}
\includegraphics[scale=0.5]{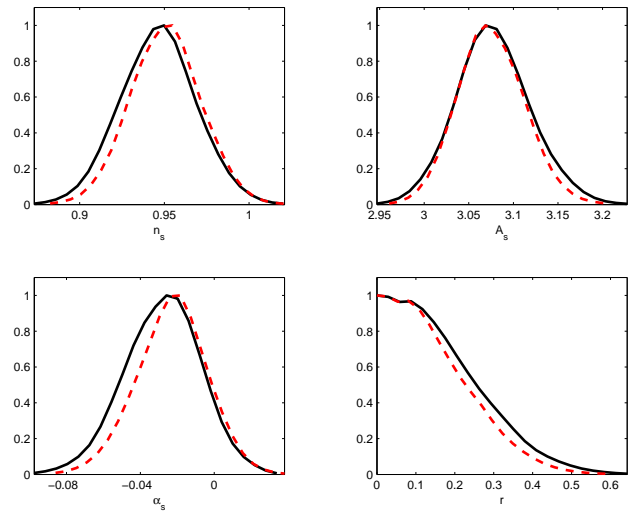}
\caption{One dimensional constraints on the inflationary parameters
$n_s$, $A_s$, $\alpha_s$ and $r$ from the current observations in
flat $\Lambda$CDM model. The black solid lines are from data sets
CMB $+$ LSS $+$ SN Ia and the red dashed lines are obtained by
taking into account the weak lensing data. \label{fig4}}
\end{center}
\end{figure}

The results are shown in Fig.\ref{fig4}. Within the $\Lambda$CDM
model, we have $n_s=0.953\pm0.0202$ ($1~\sigma$), which excludes
the scale-invariant spectrum, $n_s=1$, and the spectra with a blue tilt, $n_s>1$,
at more than $2~\sigma$ confidence level. When considering the gravitational waves, the
latest observational data yield the $95\%$ upper limit of
tensor-to-scalar ratio $r<0.346$.

%In Fig.\ref{fig4} we show the one dimensional and two
%dimensional cross correlation constraints of the inflationary
%parameters $n_s$,$\alpha_s$,$r$.

 %We find that the
%Harrison-Zel'dovich-Peebles scale-invariant (HZ) spectrum ($n_s=1$,
%$r=0$) is still disfavored more than $2~\sigma$ confidence level.
%And many hybrid inflation models and the inflation models with
%``blue" tilt ($n_s>1$) are also excluded by the current
%observations. Furthermore, the single slow-rolling scalar field with
%potential $V(\phi)\sim m^{2}\phi^{2}$, which predicts
%$(n_s,r)=(1-2/N,8/N)$, is still well within $2~\sigma$ region, while
%another single slow-rolling scalar field with potential $V(\phi)\sim
%\lambda\phi^{4}$, which predicts $(n_s,r)=(1-3/N,16/N)$, has been
%excluded more than $2~\sigma$ \cite{WMAP5GF1,KinneyWMAP5}.

We also explore the constraint on the running of the spectral index.
From all the combined data, we obtain a limit on the running of the
spectral index with $\alpha_s=-0.0231\pm0.0185$ ($1~\sigma$) for the
$\Lambda$CDM model. The error is dramatically reduced comparing with
the previous results \cite{Xiaplanck,XiaSR}, beneficial from the
more accurate observational data. No significant evidence for large
runnings of the spectral index is found. Our results also show that
the current lensing data do not improve the constraints on these
parameters significantly.

%Summary===============================================================

\section{Summary}
\label{Sum}

In this paper we have studied the constraints on the cosmological
parameters from the latest observational data including CMB, LSS, SN
Ia and weak lensing effects. We have paid particular attention to
the additional contributions from the 100 square degree cosmic shear
data. Our results show that the current weak lensing data do help to
improve the constraints on $\Omega_m$, $\sigma_8$, the total
neutrino mass $\Sigma{m_{\nu}}$, and the constant EoS of dark
energy. For other parameters, they do not add too much value due to
their large statistical errors.

%Acknowledgments=======================================================

\section*{Acknowledgements}

We acknowledge the use of the Legacy Archive for Microwave
Background Data Analysis (LAMBDA). Support for LAMBDA is provided by
the NASA Office of Space Science. We have performed our numerical
analysis on the Shanghai Supercomputer Center (SSC). We thank Yi-Fu
Cai and Jonathan Benjamin for helpful discussions. This work is
supported in part by National Science Foundation of China under
Grant Nos. 10803001, 10533010 and 10675136, and the 973 program No.
2007CB815401, and by the Chinese Academy of Science under Grant No.
KJCX3-SYW-N2.

%End===================================================================

\end{document}